\documentclass[]{aa} 
\usepackage{graphicx}
\usepackage{txfonts}
\usepackage{natbib}
\begin{document}
\title{Starspot activity and rotation of the planet-hosting star Kepler-17}
  
\authorrunning{A. S. Bonomo \& A. F. Lanza }

\titlerunning{Activity and rotation of Kepler-17}

   \author{A.~S.~Bonomo\inst{1} \and A.~F.~Lanza\inst{2} }

   \institute{INAF-Osservatorio Astrofisico di Torino, via Osservatorio, 20, 10025, Pino Torinese, Italy\\
   \email{bonomo@oato.inaf.it}
\and   
   INAF-Osservatorio Astrofisico di Catania, Via S.~Sofia, 78, 95123 Catania, Italy\\
              \email{nuccio.lanza@oact.inaf.it}
    }

   \date{Received 12 July 2012 / Accepted 20 September 2012}

 
  \abstract
   {Kepler-17 is a G2V sun-like star accompanied by a transiting planet with a mass of  $\approx 2.5$ Jupiter masses and an orbital period of  $1.486$~d, recently discovered by the \emph{Kepler} space telescope. This star is highly interesting as a young solar analogue. It is also a good candidate for a test of the tidal theories for solar-like stars.  }
   {We used about 500 days of high-precision, high-duty-cycle optical photometry collected by \emph{Kepler} to study the rotation of the star and the evolution of its photospheric active regions.}
   {We applied a maximum-entropy light curve inversion technique to model the flux rotational modulation induced by  active regions that consist of dark spots and bright solar-like faculae with a fixed area ratio. Their configuration was varied after a fixed time interval  to take  their evolution into account. Active regions were used as tracers to study stellar differential rotation, and planetary occultations were used to constrain the latitude of some spots. }
   {Our modelling approach reproduces the light variations of Kepler-17 with a standard deviation of the residuals comparable with the precision of \emph{Kepler} photometry. We find several active longitudes where individual active regions appear, evolve, and decay with lifetimes comparable to those observed in the Sun, although the star has a spotted area $\approx 10-15$ times larger than the Sun at the maximum of the 11-yr cycle. Kepler-17 shows a solar-like latitudinal differential rotation, but the fast spot evolution prevents a precise determination of its amplitude. Moveover, the stars shows a cyclic variation of the starspot area with a period of $47.1 \pm 4.5$~d, particularly evident during the last 200 days of the observations, similar to the solar Rieger cycles.  
 Possible effects of the close-in massive planet on stellar photospheric activity cannot be excluded, but require a long-term monitoring to be unambiguously detected.  }
{}

\keywords{stars: late-type -- stars: activity -- stars: rotation -- techniques: photometry -- planetary systems -- stars: individual \mbox{(Kepler-17)}}

   \maketitle
%

\section{Introduction}
The \emph{Kepler} space mission has recently announced the discovery of the 
hot Jupiter Kepler-17b. It orbits a G2V star with $K_{p}\footnote{\emph{Kepler} magnitude (cf. Sect.~\ref{observations}).}=14.14$ 
in $P=1.486$~d \citep[][ hereafter D11]{Desertetal11}. 
Thanks to additional spectra acquired 
with the SOPHIE spectrograph at the Observatoire de Haute Provence, \citet{Bonomoetal12} (hereafter B12) slightly refined the characterization of 
the Kepler-17 planetary system. In particular, they found the host star to be slightly 
hotter and younger than D11, with an effective temperature of $5781 \pm 85$~K and an 
age $< 1.8$~Gyr. 

The hosting star Kepler-17 is remarkably active, as shown 
by the out-of-transit variations 
in the \emph{Kepler} light curve with a peak-to-peak amplitude of $\approx 4\%$. 
They are produced by the rotational modulation of active regions, 
starspots and faculae, on the stellar disc.
By means of a periodogram analysis of the light curve, D11 estimated the stellar rotation 
period to be $P_{\rm rot}=11.89 \pm 0.15$~d. Intriguingly, this value of $P_{\rm rot}$ is an 
integer multiple of the orbital period, specifically eight times. 
This could be just a coincidence or the result of a 
star-planet interaction, as pointed out by D11. 

In addition, the planet occults starspots during transits, giving rise
to the typical bumps observed in the bottom of the transit profiles \citep[e.g., ][]{Silvavalioetal10,SilvavalioLanza11}.  
The integer ratio between $P_{\rm rot}$ and $P$ allowed D11 
to see a ``stroboscopic" effect with the short-cadence \emph{Kepler} data:
the spots are ``mapped" by the planet each 45~\ensuremath{^\circ} in longitude. 
The study of these spot-crossing events reveals that the planet's orbit 
is prograde and the projected spin-orbit angle is smaller than $10^{\circ}-15^{\circ}$ (see D11 for more details).

Kepler-17 is therefore an excellent candidate to study the magnetic
activity in a solar-like star younger than the Sun, through the modelling of its out-of-transit
flux variations. A similar analysis has been performed for several CoRoT planet-hosting stars. 
One of them is CoRoT-2, an active G7V star with a rotation period of 4.5~d 
and an age younger than 0.5~Gyr \citep{Alonsoetal08}. 
The spot-modelling of its light curve revealed two active longitudes and 
a cyclic oscillation of the total spotted area with a period of $28.3 \pm 4.3$~d, 
similarly to the solar Rieger cycles \citep{Lanzaetal09a}. 
Investigating the magnetic activity of G dwarfs with different ages would in principle allow one to study the Sun in time. Recently, \citet{Frascaetal11} and \citet{Frohlichetal12} have applied a spot-modelling to \emph{Kepler} photometric time series to study three young G- and K-type dwarf stars whose ages range between $\approx 50$ and $\approx 200$~Myr, deriving information on their spot evolution and surface differential rotation.

Owing to the faintness and low projected rotation velocity of Kepler-17 ($ v \sin i \approx 5$~km~s$^{-1}$), an 
investigation of its magnetic activity through Doppler imaging 
is not feasible. Therefore, modelling the stellar variability and 
studying the distortions of the transit profile when the planet occults starspots
are the only techniques that allow us to derive active longitudes where
spots preferentially form, determine the lifetime of active regions and 
activity complexes, estimate a minimum amplitude for the stellar 
differential rotation, and possibly discover short-term and/or 
long-term activity cycles, depending on the length of the 
time series.

\section{Observations}
\label{observations}
The \emph{Kepler} space telescope has an aperture of 95~cm and is designed to yield nearly continuous, high-precision photometry in the passband $423-897$~nm for $\approx 150, 000$ stars in a fixed field of view to search for planetary transits \citep[see, e.g., ][]{Boruckietal10,Kochetal10}. It orbits the Sun on an Earth-trailing orbit and, to keep the solar arrays pointed towards the Sun, 
a re-orientation of the spacecraft is required every $\approx 90$ days, a time interval called ``a quarter" in the \emph{Kepler} jargon. 
The re-orientation of the telescope produces an offset in the photon counting for a given star because its photometric mask 
is re-defined on a different CCD in the focal plane.

The light curve of Kepler-17, publicly available at the 
MAST archive\footnote{http://archive.stsci.edu/kepler/data\_search/search.php},
covers more than fifteen months of photometric
measurements, from 2009 May 13 to 2010 August 23. 
Observations are distributed along the six quarters Q1-Q6.
The raw data with the long-cadence temporal sampling, 
i.e. one point each 29.42~min, was used for our work. 
Short-cadence data (one point per minute) available for the last 
three quarters (Q4-Q6) are not particularly 
useful for our purpose because we are interested in modelling 
the out-of-transit variations on the timescale of stellar rotation or longer. 

Because of the flux offsets between adjacent quarters and the long-term instrumental trends within 
each quarter  (see \citealt{Jenkinsetal10a} and Fig.~1 in D11 for our specific case), 
the light curves corresponding to different quarters were separately treated as follows. 
First, planetary transits were removed 
from each quarter. The flux contamination 
due to starfield crowding, as estimated by the \emph{Kepler} team, 
was subtracted from the median value of the flux. Steep variations 
after the safe modes \citep{Jenkinsetal10b} were removed and 
long-term trends of clear instrumental origin were corrected by 
fitting a parabola. Lastly, the flux of each quarter 
was normalized to its median value 
then nearly matched the endpoints of adjacent quarters. 
The final light curve obtained by 
combining the six quarters contains 20\,924 data points and 
is shown in Sect.~\ref{light_curve_model} (see Fig.~\ref{lc_bestfit}). 
The median of the errors of the single photometric 
measurements is $2.18 \times 10^{-4}$ in relative flux units. 


\section{Spot modelling of wide-band light curves}
\label{spotmodel}

Reconstructing the surface brightness distribution from the rotational modulation of the stellar flux is an ill-posed problem, because the variation of the flux vs. rotational phase contains information only on the distribution of the brightness inhomogeneities vs. longitude. The integration over the stellar disc effectively cancels any latitudinal information, particularly when the inclination of the rotation axis along the line of sight is close to $90^{\circ}$, as in the present case \citep[see Sect.~\ref{model_param} and ][]{Lanzaetal09a}. Therefore, we need to include a priori information in the light curve inversion process to obtain a unique and stable map. This is  done by computing a maximum entropy (hereinafter ME) map, which has been proven to  best reproduce active region distribution and area variations for the Sun \citep[cf. ][]{Lanzaetal07}. For a different modelling approach, based on discrete starspots, see, e.g., \citet{Mosseretal09}. 
A comparison with other modelling approaches is given in, e.g., \citet{Frohlichetal09,Huberetal10,SilvavalioLanza11}.

In the present model, the stellar surface is subdivided into  elements, i.e., into 200  squares of side $18^{\circ}$, with  each element containing unperturbed photosphere, dark spots, and facular areas. The fraction of the $k$-th  element covered by dark spots is indicated by its filling factor $f_{k}$,  the fractional  area of the faculae is $Qf_{k}$, and the fractional area of the unperturbed photosphere is $1-(Q+1)f_{k}$. 
The contribution to the stellar flux coming from the $k$-th surface element at time $t_{j}$, where $j=1,..., N$  is an index numbering the $N$  points along the light curve, is given by
\begin{eqnarray}
\Delta F_{kj} & = & I_{0}(\mu_{kj}) \left\{ 1-(Q+1)f_{k} + c_{\rm s} f_{k} +  \right. \nonumber \\
  & & \left.  Q f_{k} [1+c_{\rm f} (1 -\mu_{kj})] \right\} A_{k} \mu_{kj} {w}(\mu_{kj}),
\label{delta_flux}
\end{eqnarray}
where $I_{0}$ is the specific intensity in the continuum of the unperturbed photosphere at the isophotal wavelength of the observations, $c_{\rm s}$ and $c_{\rm f}$ are the spot and facular contrasts, respectively \citep[cf. ][]{Lanzaetal04}, $A_{k}$ is the area of the $k$-th surface element,
\begin{equation}
 {w} (\mu_{kj}) = \left\{ \begin{array}{ll} 
                      1  & \mbox{if $\mu_{kj} \geq 0$}  \\
                      0 & \mbox{if $\mu_{kj} < 0$ }
                              \end{array} \right. 
\end{equation}
is its visibility, and 
\begin{equation}
\mu_{kj} \equiv \cos \psi_{kj} = \sin i \sin \theta_{k} \cos [\ell_{k} + \Omega (t_{j}-t_{0})] + \cos i \cos \theta_{k},
\label{mu}
\end{equation}
is the cosine of the angle $\psi_{kj}$ between the normal to the surface element and the direction of the observer, with $i$ being the inclination of the stellar rotation axis to the line of sight, $\theta_{k}$ the colatitude and $\ell_{k}$ the longitude of the $k$-th surface element, $\Omega$ the angular velocity of rotation of the star ($\Omega \equiv 2 \pi / P_{\rm rot}$),  and $t_{0}$ the initial time. The specific intensity in the continuum varies according to a quadratic limb-darkening law, as adopted by \citet{Lanzaetal03} for the Sun, viz. $I_{0} \propto a_{\rm p} + b_{\rm p} \mu + c_{\rm p} \mu^{2}$. The stellar flux computed at time $t_{j}$ is then: $F(t_{j}) = \sum_{k} \Delta F_{kj}$. To warrant a relative precision of about $10^{-5}$ in the computation of the flux $F$, each surface element is further subdivided into elements of $1^{\circ} \times 1^{\circ}$ and their contributions, calculated according to Eq.~(\ref{delta_flux}), are summed up at each given time to compute the contribution of the $18^{\circ} \times 18^{\circ}$ surface element to which they belong.  

We fitted the light curve by changing the value of the spot-filling factor $f$ over the surface of the star while $Q$ was held constant. Even fixing the rotation period, the inclination, and the spot and facular contrasts \citep[see ][ for details]{Lanzaetal07}, the model has 200 free parameters and suffers from  non-uniqueness and instability. To find a unique and stable spot map, we applied ME regularization, as described in \citet{Lanzaetal07}, by minimizing a functional $Z$, which is a linear combination of the $\chi^{2}$ and  the entropy functional $S$; i.e.,
\begin{equation}
Z = \chi^{2} ({\vec f}) - \lambda S ({\vec f}),
\end{equation}
where ${\vec f}$ is the vector of the filling factors of the surface elements, $\lambda > 0$   a Lagrangian multiplier determining the trade-off between light curve fitting and regularization, and { the expression of $S$ is
\begin{equation}
S = -\sum_{k} w_{k} \left[ f_{k} \log \frac{f_{k}}{m} + (1-f_{k}) \log \frac{1-f_{\rm k}}{1-m} \right],
\end{equation}
where $w_{k}$ is the relative area of the $k$-th surface element (total surface area of the star $=1$) and $m$ the default spot-covering factor that fixes the limiting values for $f_{k}$ as: $m < f_{k} < (1-m)$. In our modelling we adopted $m=10^{-6}$   \citep[cf. ][]{Lanzaetal98}. } The entropy functional $S$ is constructed in such a way that it attains its maximum value when the star is virtually immaculate ($f_{k} = m$ for every $k$). Therefore, by increasing the Lagrangian multiplier $\lambda$, the weight of $S$ in the model is increased and the area of the spots  is progressively reduced.
This gives rise to systematically negative residuals between the observations and the best-fit model when $\lambda > 0$. 

The optimal value of $\lambda$ depends on the information content of the light curve, which in turn depends on the ratio of the amplitude of its rotational modulation to the average standard deviation of its  points. In the case of Kepler-17, the  amplitude of the  rotational modulation is $\approx 0.044$, while the nominal standard deviation of the  points is $\approx 2.18 \times 10^{-4}$ in relative flux units (see Sect.~\ref{observations}), giving a signal-to-noise ratio of $\approx 200$.

To fix the optimal value of the Lagrangian multiplier $\lambda$, we compared the modulus of the mean of the residuals of the regularized best fit $|\mu_{\rm reg}|$ with the standard error of the residuals themselves, i.e., $\epsilon_{0} \equiv \sigma_{0}/\sqrt{N}$, where $\sigma_{0}$ is the standard deviation of the residuals of the unregularized best fit and $N$  the number of points in each fitted subset of the light curve having a duration  $\Delta t_{\rm f}$ (see below). We iterated until $|\mu_{\rm reg}| \simeq  \epsilon_{0}$ \citep[cf. the  case of \object{CoRoT-2} in ][]{Lanzaetal09a}. 

For the Sun, by assuming a fixed distribution of the filling factor, it is possible to obtain a good fit of the irradiance changes only for a limited time interval $\Delta t_{\rm f}$, not exceeding 14 days, which is the lifetime of the largest sunspot groups dominating the irradiance variation \citep{Lanzaetal03}. For other active stars, the value of $\Delta t_{\rm f}$ must be determined from the observations themselves, looking for the maximum data extension that allows us a good fit with the applied model (see Sect.~\ref{model_param}). 

The optimal values of the spot and facular contrasts and of the facular-to-spotted area ratio $Q$ in stellar active regions are  unknown a priori. In our model the facular contrast $c_{\rm f}$ and the parameter $Q$ enter as the product $c_{\rm f} Q$, so we can fix $c_{\rm f}$ and vary $Q$, estimating its best value 
 by minimizing the $\chi^{2}$ of the model, as shown in Sect.~\ref{model_param}. Since the number of free parameters of the ME model is large,  we used the model of \citet{Lanzaetal03}  to fix the value of $Q$. It fits  the light curve by assuming only three active regions to model the rotational modulation of the flux plus a uniformly distributed background to account for the variations of  the mean light level. This procedure is the same as adopted  for, e.g.,  \object{CoRoT-2} and \object{CoRoT-4} to fix the value of $Q$ 
\citep[cf. ][]{Lanzaetal09a,Lanzaetal09b}.  

We assumed an inclination of the rotation axis of Kepler-17 of $ i = 87.2^{\circ}$ (see Sect.~\ref{model_param}). Since the information on spot latitudes that can be extracted from the rotational modulation of the flux for such a high inclination is negligible, the ME regularization virtually puts all spots at the sub-observer latitude (i.e., $90^{\circ} -i \approx 3^{\circ}$) to minimize their area and maximize the entropy. Therefore, we are limited to mapping  only the distribution of the active regions vs. longitude, which can be achieved with a resolution of at least   $\approx 40^{\circ}-50^{\circ}$ \citep[cf. ][]{Lanzaetal07,Lanzaetal09a}. Our ignorance of the true facular contribution to the light modulation may lead  to systematic errors in the active region longitudes derived by our model, as discussed  by \citet{Lanzaetal07} for the Sun.

\section{Model parameters}
\label{model_param}

The basic stellar parameters are taken from D11 and B12 and are listed in 
Table~\ref{model_param_table}.  The limb-darkening parameters in the \emph{Kepler} bandpass were derived from the model of the transit  as discussed in D11. 

The rotation period for our spot modelling  was initially fixed at exactly eight orbital periods of the planet, following  D11, who found that there were spots occulted by the planet that reappeared at the same phase after that time interval, i.e., $11.89$~d. However, we found that the migration rates of the other spots revealed by the modelling of the out-of-transit photometry was minimized by assuming a slightly different rotation period, i.e., $12.01$~d, which we adopted for our modelling. 
\begin{table}
\noindent 
\caption{Parameters adopted for the modelling of the  light curve of Kepler-17.}
\centering
\begin{tabular}{lrr}
\hline
 & & \\
Parameter & Value & Ref.$^{a}$\\
 & & \\ 
\hline
 & &  \\
Star mass ($M_{\odot}$) & 1.16 & B12  \\
Star radius ($R_{\odot}$) & 1.05 & B12  \\
$T_{\rm eff}$ (K) & 5780 &  B12 \\
$\log g$ (cm s$^{-2}$) & 4.53 &  B12\\ 
$a_{\rm p}$ & 0.333 & BL12 \\
$b_{\rm p}$ & 0.929 & BL12 \\
$c_{\rm p}$ & -0.262 & BL12 \\ 
$P_{\rm rot}$ (days) & 12.01 & BL12 \\
$\epsilon_{\rm rot}$ & $4.66 \times 10^{-5}$ & BL12 \\ 
Inclination (deg) & 87.22 & D11  \\
$c_{\rm s}$  & 0.677 & L04 \\
$c_{\rm f}$  & 0.115 & L04 \\ 
$Q$ & 1.6  & BL12 \\ 
$\Delta t_{\rm f}$ (days) & 8.733 & BL12 \\ 
& &   \\
\hline
\label{model_param_table}
\end{tabular}
~\\
$^{a}$ References: B12: \citet{Bonomoetal12}; D11: \citet{Desertetal11}; L04: \citet{Lanzaetal04}; BL12: present study. 
\end{table}

The polar flattening of the star owing to the centrifugal potential was computed in the Roche approximation with a rotation period of 12.01~d. The relative difference between the equatorial and the polar radii is $\epsilon_{\rm rot} = 4.66 \times 10^{-5}$, which induces a completely negligible relative flux variation of $\approx 2 \times 10^{-6}$  for a spot coverage of $\approx 4$ percent, as a consequence of the gravity darkening of the equatorial regions of the star. 

The inclination of the stellar rotation axis is constrained by the observation of D11 that the light anomalies detected in successive transits are compatible with the same spots being repeatedly occulted as they move over the stellar disc owing to stellar rotation. This is possible only if the disc-projected trajectory of the spots remains inside the belt occulted by the planet. In other words, this is an indication that the sky-projected misalignment between the stellar spin and the orbital angular momentum is smaller than $\pm (10^{\circ}-15^{\circ})$ (cf. D11 for details). 
 Although the possibility that the stellar spin is not aligned with the orbital angular momentum cannot be excluded with certainty because only the sky-projected angle between the two vectors has been measured, we regard this occurrence as highly improbable and assume that the rotation axis is inclined with respect to the line of sight of the same angle as the orbit normal. Moreover, the stellar $v \sin i = 4.7 \pm 1.0$~km~s$^{-1}$ and the estimated stellar radius provide an inclination close to $90^{\circ}$ for the stellar rotation period (cf. D11).

The maximum time interval $\Delta t_{\rm f}$ that our model can accurately fit with a fixed distribution of active regions was determined by dividing the total interval, $T= 497.785$~d, into $N_{\rm f}$ equal segments, i.e., $\Delta t_{\rm f} = T/N_{\rm f}$, and looking for the minimum value of $N_{\rm f}$ that allows us a good fit of the light curve, as measured by the $\chi^{2}$ statistics. We found that for $N_{\rm f} < 57 $ the quality of the best fit degrades appreciably with respect to higher values, owing to a  substantial evolution of the pattern of surface brightness inhomogeneities. Therefore, we adopted $N_{\rm f} = 57$, so that  $\Delta t_{\rm f} = 8.733$~d is the maximum time interval to be fitted with a fixed distribution of surface active regions  to  estimate the best value of the parameter $Q$ (see below). 

To evaluate the spot contrast, we adopted the same mean temperature difference as derived for sunspot groups from their bolometric contrast, i.e. 560~K \citep{Chapmanetal94}. The effective temperature of the unspotted photosphere is $5780 \pm 85$~K, i.e., very similar to that of the Sun (cf. B12).  
In other words, we assumed a spot effective temperature of $ 5220$~K, yielding a contrast $c_{\rm s} = 0.677$ in the bolometric passband \citep[cf. ][]{Lanzaetal07}.
A different spot contrast changes the absolute spot coverages, but neither significantly affects their longitudes or their time evolution, as discussed in detail by \citet{Lanzaetal09a}. Therefore, since the spot temperature is only estimated, we neglected the difference in the contrast $c_{\rm s}$ between the \emph{Kepler} bandpass and the bolometric passband in our analysis. 

\begin{figure}[b]
\centerline{
\includegraphics[width=8cm,height=6cm,angle=0]{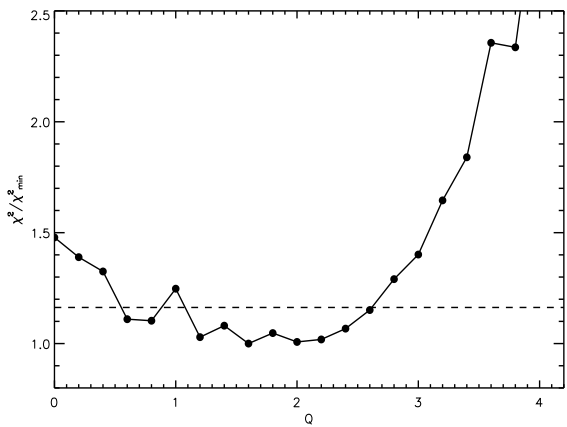}} 
\caption{
Ratio of the $\chi^{2}$ of the composite best fit of the entire time series of Kepler-17   
to its minimum value vs. the parameter $Q$, i.e., the ratio of the area of the faculae to that of the cool spots in active regions. The horizontal dashed line indicates the 95 percent confidence level for $\chi^{2}/\chi_{\rm min}^{2}$, determining the interval of acceptable $Q$ values.
}
\label{qratio}
\end{figure}

In our model, the facular contrast is assumed to be solar-like with $c_{\rm f} = 0.115$ \citep{Lanzaetal04}.    
The best value of the area ratio $Q$ between the faculae and the spots in the active regions was estimated by means of the three-spot model by \citet[][ cf. Sect.~\ref{spotmodel}]{Lanzaetal03}. In Fig.~\ref{qratio}, we plot the ratio $\chi^{2}/ \chi^{2}_{\rm min}$ of the total $\chi^{2}$ of the composite best fit of the entire time series to its minimum value $\chi^{2}_{\rm min}$, versus $Q$, and indicate the 95 percent confidence level as derived from the F-statistics \citep[e.g., ][]{Lamptonetal76}. The choice of $\Delta t_{\rm f} = 8.733$~d allows us to fit the rotational modulation of the active regions for the longest time interval during which they remain stable, modelling both the flux increase due to the facular component when an active region is close to the limb and the flux decrease due to the dark spots when the same region transits across the central meridian of the disc. In this way, a measure of the relative facular and spot contributions can be obtained, leading to an  estimate of $Q$. 
Fig.~\ref{qratio} shows that the best value is $Q=1.6$, with an acceptable range extending from~$\approx 0.6$~to~$\approx 2.6$.  Therefore, we adopted $Q=1.6$ for our modelling in Sect.~\ref{results}. We comment on the value of $Q$ in more detail in Sect.~\ref{conclusions}.

\begin{figure*}[!t]
\centerline{
\includegraphics[width=12cm,height=16cm,angle=90]{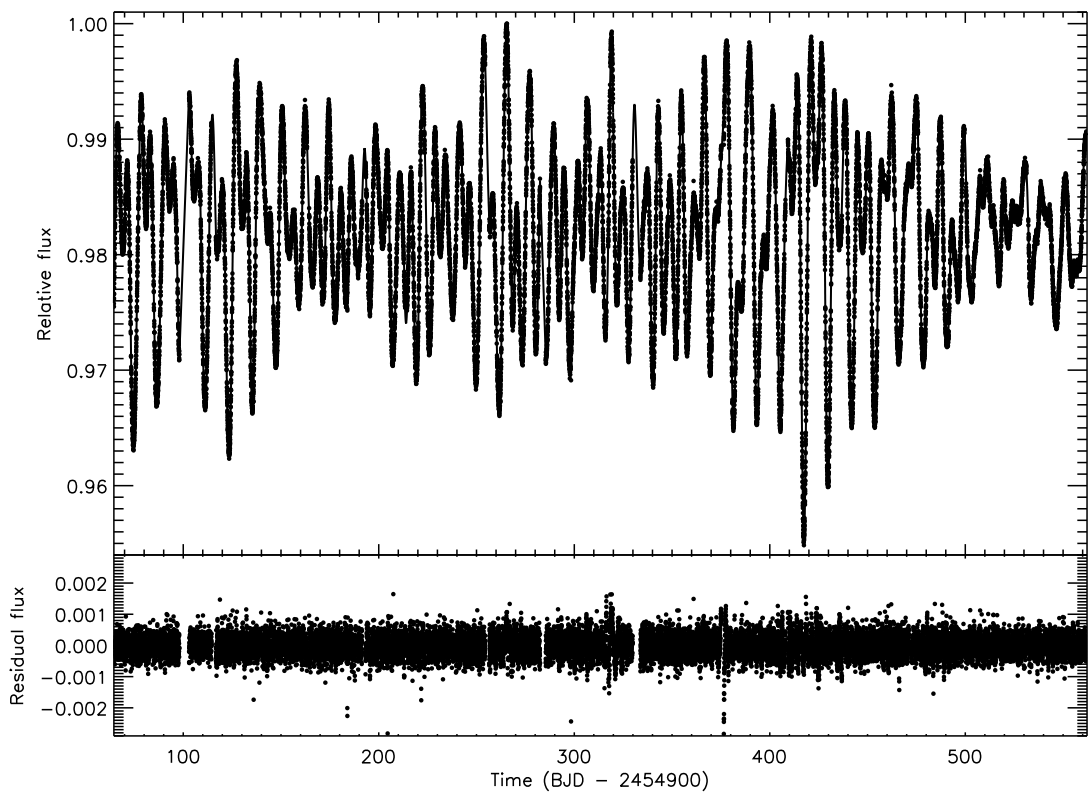}} 
\caption{{\it Upper panel:} The ME-regularized composite best fit to the out-of-transit light curve of Kepler-17 obtained for $Q=1.6$. The flux is in relative units, i.e., normalized to the maximum observed flux along the light curve.  {\it Lower panel:} The residuals from the composite best fit versus the time.
}
\label{lc_bestfit}
\end{figure*}
\begin{figure}[!b]
\centerline{
\includegraphics[width=6cm,height=8cm,angle=90]{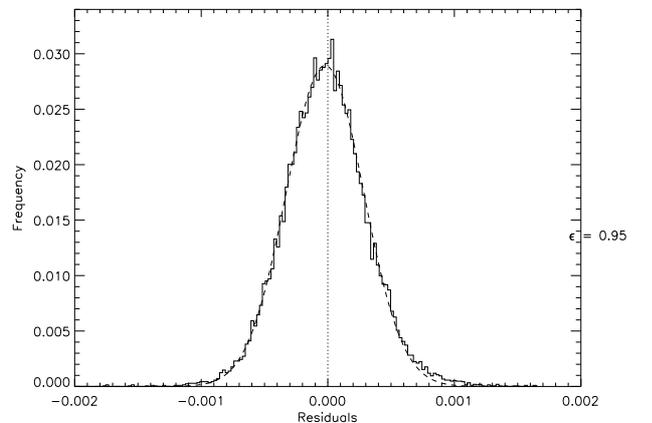}} 
\caption{Distribution of the residuals to the composite ME-regularized best fit shown in Fig.~\ref{lc_bestfit}. The dashed line is a Gaussian fit to the residual distribution, while the vertical dotted line marks the zero value of the residuals. 
}
\label{hist_residuals}
\end{figure}

\section{Results}
\label{results}

\subsection{Light curve models}
\label{light_curve_model}

We applied the model of Sect.~\ref{spotmodel} to the out-of-transit  light curve of Kepler-17, considering time  intervals  $\Delta t_{\rm f} = 8.733$~d. 
The best fit without regularization ($\lambda = 0$) has a mean  $\mu_{\rm res} = 3.845 \times 10^{-6}$ and a standard deviation of the residuals $\sigma_{0} = 3.032 \times 10^{-4}$ in relative flux units. The Lagrangian multiplier $\lambda$ is iteratively adjusted until the mean of the residuals $\mu_{\rm res} = -1.583 \times 10^{-5} \simeq \sigma_{0}/\sqrt{N}$, where $N  =  367$ is the mean number of  points in each fitted light curve interval  $\Delta t_{\rm f}$; the standard deviation of the residuals of the regularized best fit is $\sigma = 3.311 \times 10^{-4}$. 

The composite best fit to the entire light curve is shown in the upper panel of Fig.~\ref{lc_bestfit}, while the residuals are plotted in the lower panel. The best fit is always very good, with a standard deviation of the residuals  $\approx 1.52$ times the median of the errors of the 
photometric measurements as given by the \emph{Kepler} pipeline. The distribution of the residuals is plotted in Fig.~\ref{hist_residuals} and is well fitted by a Gaussian with a standard deviation of $3.056 \times 10^{-4}$ for absolute values of the residuals lower than $ \approx 6 \times 10^{-4}$ in relative flux units. For residuals greater than $\approx 6 \times 10^{-4}$ there is a remarkable asymmetry in the distribution with an excess of positive residuals.  
A periodogram of the residuals shows a highly significant peak at a period of $1.505 \pm 0.005$~d, very close to the orbital period, and several lower peaks at periods between $\approx 1.2$ and $\approx 3$~d. Putting the residuals in phase with the orbital period, we see the occultation of the planet and a possible phase-dependence of the reflected light (see B12).

\subsection{Longitude distribution of active regions and stellar differential rotation}
\label{spot_model_res}
\begin{figure*}[!t]
\centerline{
\includegraphics[width=12cm,height=18cm,angle=90]{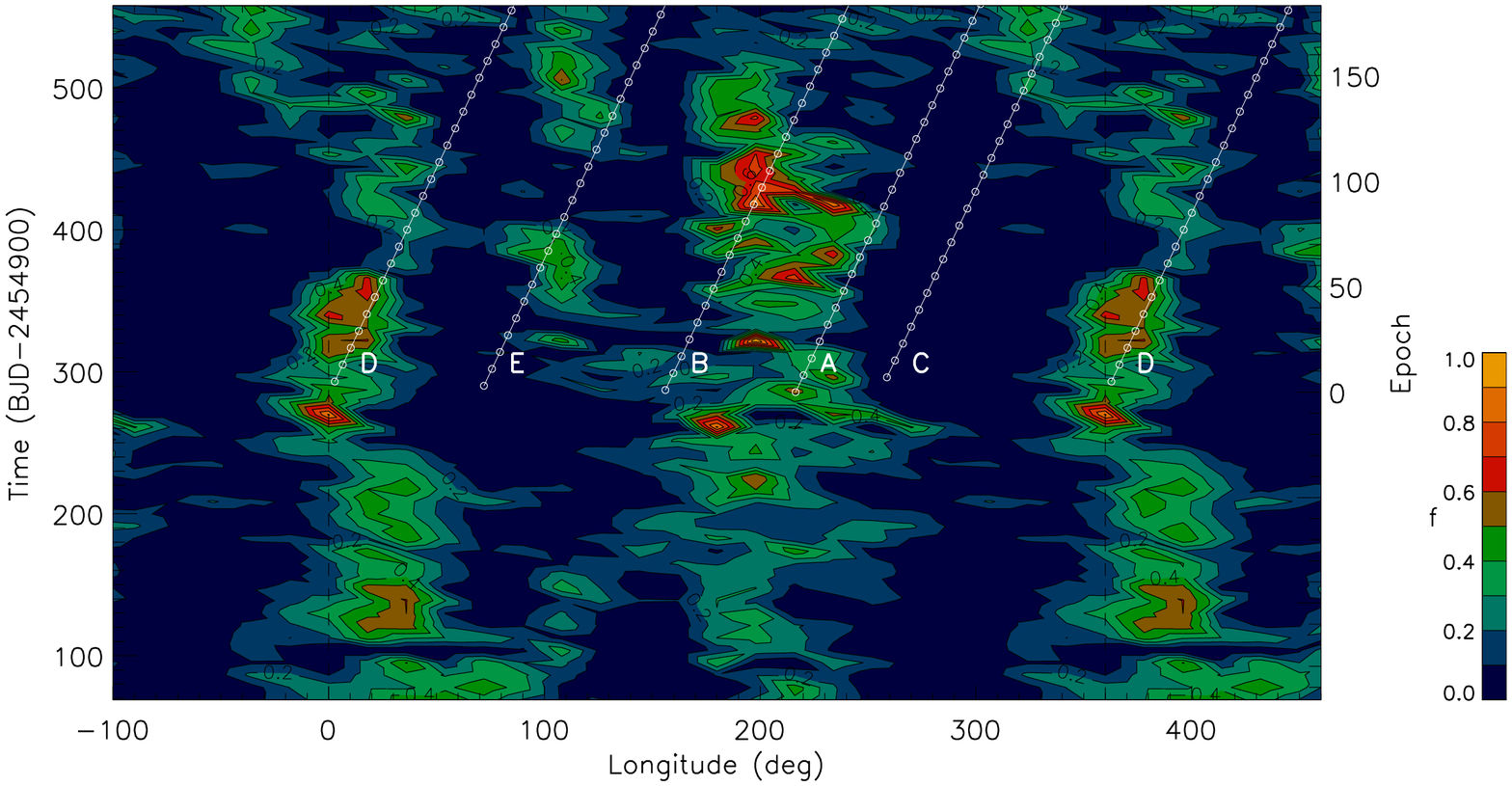}} 
\caption{Distribution of the spot-filling factor vs. longitude and time for $Q=1.6$. The values of the filling factor were normalized to their maximum $f_{\rm max} =0.01553$ with orange-yellow indicating the maximum and dark blue the minimum (see the colour scale on the right lower corner of the figure for the correspondence between the colour and the normalized filling factor).  Note that the longitude scale is extended beyond $0^{\circ}$ and $360^{\circ}$ to help following the migration of the starspots.  The tracks of the five spots occulted during the planetary transits after D11 are also reported. The open circles mark the time intervals of eight transits after which the same spots are detected again during the transits. The straight lines connecting the circles trace the migration of those spots in our reference frame; each line is labelled with the name of the corresponding spot, as indicated in Fig.~11 of D11.
}
\label{long_distr}
\end{figure*}

The distribution of the spot-filling factor $f$ vs. the longitude and the time is plotted in Fig.~\ref{long_distr}. 
 The longitude zero corresponds to the point intercepted on the stellar photosphere by the line of sight to the centre of the star  at BJD~2454964.5109, i.e., the sub-observer point at the initial epoch. The reference frame rotates with the star with a fixed period of $12.01$~d and the longitude increases in the same direction as the stellar rotation and the orbital motion of the planet. This is consistent with the reference frames adopted in our previous  studies \citep[e.g., ][]{Lanzaetal09a,Lanzaetal09b}, but does not allow a direct comparison of the mapped active regions with the dips in the light curve. 

Our map shows that the individual starspots evolve with  timescales of tens of days, which makes it difficult to trace the evolution and migration of the active regions in an unambiguous way. Nevertheless,  two main active longitudes, where individual spots form and evolve, can be identified with  confidence and appear to rotate on the whole with the rotation period of our reference frame. The migration rate of individual spots or groups of spots within each longitude is variable, which suggests that individual spots are forming at different latitudes on a differentially rotating star or, alternatively, there are several spots at close longitudes that are evolving to mimic spot migration. Unfortunately, there is no information on the starspot latitude from our mapping  technique and even the hemisphere cannot be determined because the inclination is close to $90^{\circ}$ (cf. Sect.~\ref{spotmodel}). Nevertheless, some additional information can be extracted in this particular case from the occultation of the spots by the planet during transits. Specifically, we can exploit the results of D11 to constrain the latitude of some of the starspot trails that are seen in Fig.~\ref{long_distr}. 

\citet{Desertetal11} identified five starspots that are repeatedly occulted every eight transits and labelled them A, B, C, D, and E (cf. their Figs. 11 and 12).  The transit profile distortions associated with spot C are barely visible in their Fig.~11, thus spot C is certainly significantly smaller than the other four spots. We plot in Fig.~\ref{long_distr} the migration of the starspots detected by D11. The initial epoch $E(0)$ of D11  is equal to the epoch of the first mid transit as reported in their Table~3 (D\'esert, priv. comm.) so we can easily convert their phases of maximum starspot visibility  into longitudes in our reference frame. We find a  good association between their starspots A, B, D, and E and some trails of spots found with our approach. For starspot C we find almost no coincidence, as expected owing to its smaller spotted area. The resolution of the transit-mapping method is of the order of some degrees \citep[cf., e.g., ][ and references therein]{Silvavalioetal10,SilvavalioLanza11} while that of our mapping based on the out-of-transit light curve is  approximately  $40^{\circ}-50^{\circ}$. Therefore, we are not able to resolve the occulted spots with the same detail as D11,  especially when there are several small spots close in longitude. 
Note that a detailed comparison of the spot maps based on the in-transit and out-of-transit photometry is outside the scope of the present work because it requires a detailed modelling of the transit profile distortions induced by occulted starspots \citep[see ][ for such a detailed comparison for CoRoT-2]{SilvavalioLanza11}. Here we limit ourselves to exploiting the preliminary modelling of D11 to find that  our  trails of spots with a positive migration rate, i. e., rotating faster than our reference frame, are located inside the belt that is occulted by the planet. A faster migration is particularly evident in the case of our spot trails associated with their spots B and D, at least for a significant part of the considered time interval. On the other hand,  the intrinsic evolution of the pattern makes the determination of the migration rates of our spots  associated with their spots A and E less certain, although a positive migration is suggested. On the whole, these results indicate that the occulted spots located at low latitudes are rotating faster than the overall pattern of the active longitudes, thus giving evidence of a solar-like differential rotation in Kepler-17, i.e., with the equator rotating faster than the poles.

Finally, we note that the time resolution of our spot models is not adequate to look for a possible modulation of the stellar activity with the orbital period of its  close-in massive planet, as suggested by \citet{Shkolniketal08} or \citet{Lanza11}. Given the short orbital period of the planet, a different approach should be used to search for signatures of a possible star-planet interaction using short-cadence data, as for CoRoT-2 \citep[cf., e.g., ][]{Paganoetal09}. 

\subsection{Variation of the spotted area}

The lifetimes of the active longitudes traced in Fig.~\ref{long_distr} range from $\approx 75$ to $\approx 330$ days and possibly longer, given the limited time interval covered by our observations. Individual active regions have a lifetime of about a few tens of days. This is similar to what we observe in the Sun where complexes of activity consisting of several active regions, forming approximately around the same longitude, have lifetimes up to $5-6$ months. The duration of the light anomalies observed during the transits reaches to $\approx 0.02$ in phase units, indicating a size of the largest occulted active regions of at least $\approx 40^{\circ}$ in longitude, i.e.,  $3-4$ times  that of the largest sunspot groups. 

\begin{figure}[t]
\centerline{
\includegraphics[width=6cm,height=9cm,angle=90]{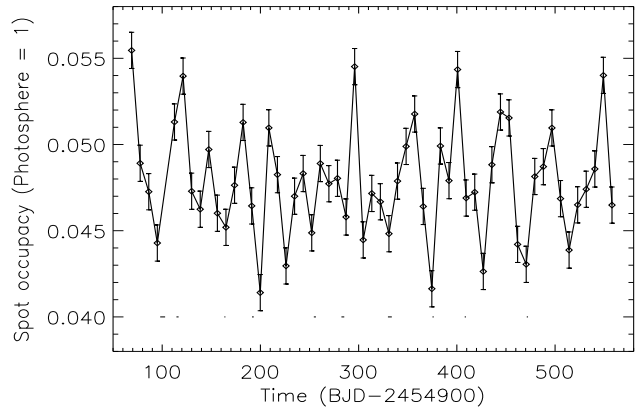}} 
\caption{Total spotted area as derived from the regularized ME models vs. time for $Q=1.6$. The lower horizontal ticks mark the gaps in the photometric time series longer than 24 hours.}
\label{total_area}
\end{figure}

The variation of the total starspot area vs. the time is plotted in Fig.~\ref{total_area}. 
The error bars have a semi-amplitude of three standard deviations as derived from the uncertainty of the photometry, but systematic errors associated with the assumptions of the data reduction and the model are  not included. Specifically, 
the variations on time scales longer than the duration of one quarter, i.e., $90-100$ days, cannot be reconstructed from \emph{Kepler} photometry because of the flux jumps from one quarter to the next (see Sect.~\ref{observations}). Therefore, we are limited to study the  variations on timescales shorter than three months. Moreover, the presence of gaps in the observations can introduce systematic errors in the measurement of the spotted area. This is the case of the gap beginning at BJD~2454997.982 for a duration of 5.0268~d. Since the time interval adopted for our individual best fits is  $\Delta t_{\rm f}= 8.733$~d, this loss of data implies a systematically lower value of the total spotted area because the ME regularization removes spots at the longitudes not constrained by the observations. Fortunately, the other gaps in the photometric time series are much shorter than $\Delta t_{\rm f}$, thus no other value of the area is significantly affected. To show the distribution of the gaps, we plot line segments with a length equal to their duration at the level 0.04 in Fig.~\ref{total_area} considering all interruptions with a duration longer than 24 hours. In the analysis presented below, we discarded the area value at BJD 2455003.81 because it is affected by the gap that started at 245997.982. 

\begin{figure}[t]
\centerline{
\includegraphics[width=6cm,height=8cm,angle=90]{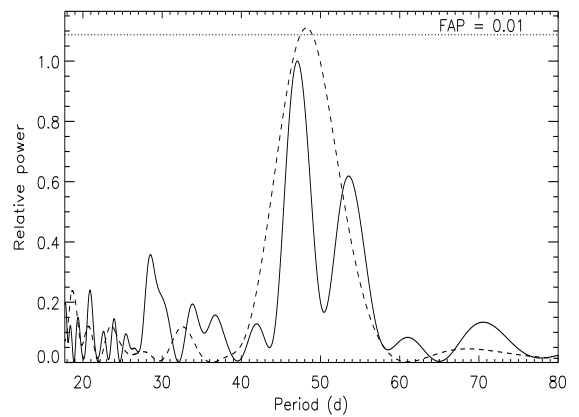}} 
\caption{Lomb-Scargle periodogram of the variation of the spotted area. The solid line gives the normalized power vs. the period for the whole time interval, while the dashed line gives the power  for the time interval from BJD  2455230.8696 to BJD 2455457.9294 with the same normalization as adopted for the periodogram of the whole interval. The horizontal dotted line marks the 99 percent confidence level (FAP = 0.01). 
}
\label{pow_total_area}
\end{figure}

The Lomb-Scargle periodogram of the entire time series of the area values is plotted in Fig.~\ref{pow_total_area} (solid line) together with the power level corresponding to a false-alarm probability of 0.01 as evaluated according to \citet{HorneBaliunas86}. The main peak corresponds to a period of $47.1 \pm 4.5$~d and has a false-alarm probability (FAP) of $2.0$ percent. The value of the FAP was confirmed  by performing an analysis of 50,000 random Gaussian noise time series with the same  sampling as our area data series.
We also plot the periodogram of the time interval from BJD~2455230.869 to 2455457.929 (dashed line) because we see a regular oscillation of the total spotted area with a period of $\approx 48$~d during that  interval in  Fig.~\ref{total_area}. The main periodogram peak corresponds to a period of $48.2 \pm 9.0$~d with an FAP of $0.44$ percent. 

\begin{figure}[b]
\centerline{
\includegraphics[width=6cm,height=8cm,angle=90]{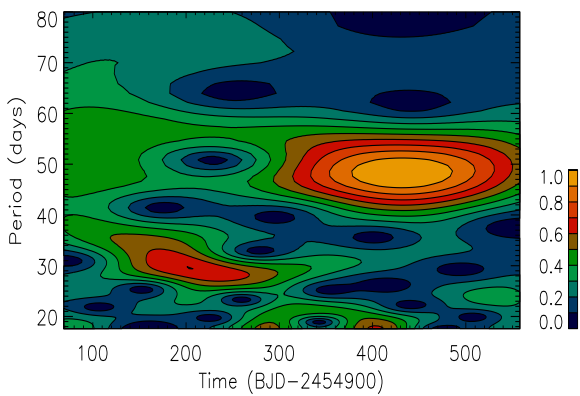}} 
\caption{Amplitude of the Morlet wavelet of the total spotted area variation vs. the period and the time. The amplitude was normalized to its maximum value. Different colours indicate different relative amplitudes from the maximum (yellow) to the minimum (dark blue) as indicated in the colour scale in the right lower corner.
}
\label{wavelet_total_area}
\end{figure}

The time variation of the frequency of the spotted area modulation is best represented by means of a wavelet amplitude. We plot in 
Fig.~\ref{wavelet_total_area} the amplitude of the Morlet wavelet versus the period and the time \citep[see, e.g., ][ for details]{HempelmannDonahue97}. The wavelet parameters are adjusted to give a time resolution of $\approx 100$~d for a period of about $48$~d and a relative period resolution of $\Delta P / P \approx 0.06$. We see that in the initial part of the dataset  there is a periodicity of $\approx 30$~d that corresponds to secondary peaks  in the periodogram of the whole time series  whose false-alarm probability is $ > 50$ percent (cf. Fig.~\ref{pow_total_area}, where the frequency resolution is better than in the case of the wavelet). On the other hand, during the second half of the time series we see a clear periodicity of $\approx 50$~d, which corresponds to the significant peak in the periodogram. We conclude that the total spotted area of Kepler-17 showed an oscillation with a period of $ 47.1 \pm 4.5$~d. This behaviour is reminiscent of the short-term oscillations of the total sunspot area found close to the maxima of some of the 11-year solar cycles. They were called Rieger cycles because they were first detected in the periodicity of occurrence of large solar flares by \citet{Riegeretal84}. In the Sun, the periodicity is $\approx 160$~d with small variations from one cycle to the other, although only some of the sunspot maxima show evidence of this short-term periodicity \citep{Oliveretal98,KrivovaSolanki02,Zaqarashvilietal10}. A behaviour similar to that of Kepler-17 was found in CoRoT-2, a G7V star that showed a Rieger-type cycle in the variation of its spotted area with a period of $28.9 \pm 4.3$~d \citep{Lanzaetal09a}.


\section{Discussion and conclusions}
\label{conclusions}

The application of a spot model to reproduce the optical light curve of Kepler-17 shows that the spot pattern is almost stable for a timescale of $\approx 9$~d because the residuals to our best fits have a standard deviation of $\sigma \approx 3.3 \times 10^{-4}$ in relative flux units that is only $\approx 50$ percent greater than the mean error attributed to the photometric measurements by the \emph{Kepler} pipeline. Our facular-to-spotted area ratio  $Q=1.6$  is significantly lower than the value $Q_{\odot}=9.0$ adopted for the modelling of the light curves of the Sun-as-a-star  by \citet{Lanzaetal07}. However, this lower value of $Q$ is typical of sun-like stars that are more active than the Sun \citep[cf. ][]{Lanzaetal09a,Lanzaetal10,Lanzaetal11a,Lanzaetal11b} and suggests an increasing weight  of the dark spots in the photometric variations as a star becomes more active, as indicated by the results of, e.g., \citet{Radicketal98} and \citet{Lockwoodetal07}.

Our models found several active longitudes that clustered on opposite hemispheres with a separation of $\approx 180^{\circ}$ for more than half the observation interval. This explains the two peaks found in the periodogram of the light curve by D11, one corresponding to the rotation period and the other to its first harmonic. 

\citet{Bonomoetal12} suggested an age younger than 
$1.8$~Gyr, while D11 derived an age of $3.0 \pm 1.6$~Gyr for Kepler-17. As noted by B12, the age determined by means of standard gyrochronology is only $0.9 \pm 0.2$~Gyr, while considering the effects of the close-in planet on the evolution of  the stellar angular momentum \citep{Lanza10}, an age of $1.7 \pm 0.3$~Gyr is estimated  that seems more compatible with the age found from isochrone fitting. 

Since the rotation period of the star is longer than the orbital period, tides remove angular momentum from the orbit to spin up the star and lead to  orbital decay. The timescale for the engulfment of the planet can be estimated according to \citet{OgilvieLin07} as 
$\tau_{a} \simeq 0.048 (Q^{\prime}_{\rm s}/10^{6})$~Gyr, where $Q^{\prime}_{\rm s}$ is the modified tidal quality factor of the star. This timescale is much shorter than the lifetime of the system on the main sequence if we adopt  $Q^{\prime}_{\rm s} \approx 10^{6}$, i.e., the value derived from the observed circularization periods of close binary systems in open clusters of different ages. Together with the observations of several other stars with  massive planets on very tight orbits, this suggests that $Q^{\prime}_{\rm s}$ should be much higher (i.e., the tidal dissipation much lower) in those star-planet systems than in close binary systems that consist of two main-sequence stars. The difference in the $Q^{\prime}_{\rm s}$ value could arise because the stellar  rotation is  far from being synchronized with the orbital motion of its planet. In this case, considering the dissipation of the tides inside the convection zone, \citet{OgilvieLin07} predicted $Q^{\prime}_{\rm s} \ga 5 \times 10^{9}$, which implies an infall timescale longer than the system lifetime.

A lower limit on the value of $Q^{\prime}_{\rm s}$ can be set by an accurate timing of the transits over a time interval of a few decades because for $Q^{\prime}_{\rm s}= 10^{6}$ we expect a variation of the orbital period of $\Delta P_{\rm orb}/P_{\rm orb} \approx 5 \times 10^{-8}$ in ten years. It produces a variation of  $\approx 8$~s in the epoch of the mid transit in 10 years. The accuracy reported by D11 is  $\pm \, 2.4$~s for their initial transit epoch, implying that a $Q_{\rm s}^{\prime}$ of about $10^{6}$ should give an orbital period acceleration detectable in a few years with a space-borne photometer. The recently approved  extension of the \emph{Kepler} mission till 2016 is therefore an interesting opportunity to perform such measurements. A model of the light perturbations that are caused by the spots occulted during the transits may possibly be used to improve the accuracy since D11 found small $O-C$ timing oscillations with a period of approximately half the stellar rotation period, i.e., likely associated with starspots on opposite stellar hemispheres (cf. their Fig.~10). 

We can estimate an approximate lower limit to the differential rotation of Kepler-17 finding $\Delta \Omega/\Omega \approx 0.10-0.16$ from the migration rates of the different trails of spots as seen  in Fig.~\ref{long_distr}. Given the rapid evolution of the individual spots, this value is not only approximate, but  depends critically on the way individual spot trails are identified. Therefore, we caution to regard  this estimate as remarkably uncertain (cf. Sect.~\ref{spot_model_res}). For comparison, in k$^{1}$~Ceti, a dwarf star with a spectral type G5V and a mean rotation period of $\approx 9$~d, \citet{Walkeretal07} estimated a relative differential rotation amplitude of $\approx 0.090 \pm 0.006$ by modelling the optical photometry obtained by the satellite MOST. A result similar to that of Kepler-17 was found by \citet{Crolletal06} for $\epsilon$~Eridani, a K2V star with a rotation period of $\approx 11.4$~d, showing a relative amplitude of the differential rotation of $0.11 \pm 0.03$. On the other hand, the more rapidly rotating star CoRoT-2 ($P_{\rm rot} \approx 4.52$~d) shows very little surface differential rotation with a relative amplitude smaller than $\approx 1$ percent \citep{Lanzaetal09a}. 
In the young ($\approx 50$~Myr) K2V star KIC~8429280, \citet{Frascaetal11} found $\Delta \Omega/ \Omega \approx 0.05$ while for the G2V stars KIC~7985370 and KIC~7765135, which have an age of $100-200$~Myr, \citet{Frohlichetal12} found $\Delta \Omega / \Omega \approx 0.07-0.08$.
The differences  can be attributed  to the dependence of the differential rotation amplitude on the stellar rotation period and effective temperature as well as to the limited range of latitudes covered by starspots that can vary  along the activity cycle. Moreover, the amplitude of the differential rotation could change along the activity cycle in rapidly rotating, highly active stars, such as the above mentioned \emph{Kepler} targets that have $P_{\rm rot}$ ranging from 1.2 and 2.8~d \citep[cf. ][]{Frascaetal11,Frohlichetal12}. Therefore, individual results coming from spot modelling can be compared only in a statistical sense. Considering the general trend found by \citet{Barnesetal05}, the amplitude of the differential rotation estimated for Kepler-17 appears within the expected range for a star with its rotation period and effective temperature.  Nevertheless, a more extended series of data is required to derive a firm conclusion on this point as well as on a possible influence of the planet on spot activity as conjectured by, e.g.,  \citet{Lanza08,Lanza11}.

An interesting result of our analysis is the short-term spot cycle with a period of $47.1 \pm 4.5$~d, which is clearly detected in the second half of the data series. This phenomenon is reminiscent of the solar Rieger cycles because of its timescale and transient nature. The periodicity of the spotted area variations is close to $4$ times the mean rotation period of the star, while in the  Sun it is $\approx 6$ times and in  CoRoT-2 is $\approx 6.5$ times the rotation period. Those oscillations of the spotted area may be associated with hydromagnetic Rossby-type waves propagating in the upper part of the convection zone or at the interface between the radiative interior and the convection zone where the dynamo is probably located \citep{Lou00,Zaqarashvilietal10}. Since only a few examples of stars displaying Rieger-type cycles are known \citep[cf., ][]{Massietal05,Lanzaetal09a}, the new results on Kepler-17 are  particularly interesting for a better understanding of this phenomenon in the framework of the solar-stellar connection.

As new \emph{Kepler} data become available, it will be possible to refine the conclusions of the present study by investigating a longer time span. This is needed to assess the duration and the frequency of the Rieger-type oscillations in the total starspot area as well as to derive a definite conclusion on the possible impact of the planet on stellar activity.

\begin{acknowledgements}
The authors wish to thank an anonymous referee for several valuable comments, the NASA and the \emph{Kepler} team for 
giving public access to the \emph{Kepler} data.
A.~S.~Bonomo gratefully acknowledges support through an INAF/HARPS-N fellowship.
Active star research and exoplanetary studies at INAF-Osservatorio Astrofisico di Catania 
and Dipartimento di Fisica e Astronomia dell'Universit\`a degli Studi di Catania 
are funded by MIUR ({\it Ministero dell'Istruzione, dell'Universit\`a e della Ricerca}) and by {\it Regione Siciliana}, whose financial support is gratefully
acknowledged. 
This research has made use of  the ADS-CDS databases, operated at the CDS, Strasbourg, France.
\end{acknowledgements}


\begin{thebibliography}{}
\bibliographystyle{aa}

\bibitem[\protect\citeauthoryear{Alonso et al.} {2008}]{Alonsoetal08}
Alonso, R., Auvergne, M., Baglin, A., et al. 2008, \aap, 482, L21


\bibitem[Barnes et al. (2005)]{Barnesetal05} 
Barnes, J.~R., Collier Cameron, A., Donati, J.-F., et al.\ 2005, \mnras, 357, L1 

\bibitem[Bonomo et al. (2012)]{Bonomoetal12}
Bonomo, A. S., H\'ebrard, G., Santerne, A., et al.\ 2012, \aap, 538, A96 (B12)

\bibitem[Borucki et al.(2010)]{Boruckietal10} 
Borucki, W.~J., Koch, D., Basri, G., et al.\ 2010, Science, 327, 977 

\bibitem[\protect\citeauthoryear{Chapman et al.}{1994}]{Chapmanetal94}
Chapman, G. A., Cookson, A. M., Dobias, J. J. 1994, \apj, 432, 403

\bibitem[Croll et al.(2006)]{Crolletal06} 
Croll, B., Walker, G.~A.~H., Kuschnig, R., et al.\ 2006, \apj, 648, 607 

\bibitem[\protect\citeauthoryear{D\'esert et al.}{2011}]{Desertetal11}
D\'esert, J.-M., Charbonneau, D., Demory, B.-O., et al.\  2011, \apjs, 197,14 (D11)


\bibitem[Frasca et al. (2011)]{Frascaetal11} 
Frasca, A., Fr{\"o}hlich, H.-E., Bonanno, A., et al.\ 2011, \aap, 532, A81

\bibitem[Fr{\"o}hlich et al.(2009)]{Frohlichetal09} 
Fr{\"o}hlich, H.-E., K{\"u}ker, M., Hatzes, A.~P., \& Strassmeier, K.~G.\ 2009, \aap, 506, 263 


\bibitem[Fr{\"o}hlich et al. (2012)]{Frohlichetal12} 
Fr{\"o}hlich, H.-E., Frasca, A., Catanzaro, G., et al.\ 2012, \aap, in press {\tt [arXiv:1205.5721]}

\bibitem[Hempelmann \& Donahue (1997)]{HempelmannDonahue97} Hempelmann, A., \& Donahue, R.~A.\ 1997, \aap, 322, 835 


\bibitem[Horne \& Baliunas (1986)]{HorneBaliunas86} 
Horne, J.~H., \& Baliunas, S.~L.\ 1986, \apj, 302, 757

\bibitem[Huber et al.(2010)]{Huberetal10} 
Huber, K.~F., Czesla, S., Wolter, U., \& Schmitt, J.~H.~M.~M.\ 2010, \aap, 514, A39


\bibitem[\protect\citeauthoryear{Jenkins et al.} {2010a}]{Jenkinsetal10a} 
Jenkins, J. M., Caldwell, D. A., Chandrasekaran, H. et al. 2010a, \apj, 713, L120


\bibitem[\protect\citeauthoryear{Jenkins et al.} {2010b}]{Jenkinsetal10b} 
Jenkins, J. M., Caldwell, D. A., Chandrasekaran, H. et al. 2010b, \apj, 713, L87


\bibitem[\protect\citeauthoryear{Koch et al.} {2010}]{Kochetal10} 
Koch, D.~G., Borucki, W.~J., Basri, G., et al. 2010, \apj, 713, L79 


\bibitem[Krivova \& Solanki (2002)]{KrivovaSolanki02} 
Krivova, N.~A., \& Solanki, S.~K.\ 2002, \aap, 394, 701 


\bibitem[\protect\citeauthoryear{Kurucz}{2000}]{Kurucz00}
Kurucz, R. L. 2000: http://kurucz.harvard.edu/



\bibitem[\protect\citeauthoryear{Lampton et al.}{1976}]{Lamptonetal76}
Lampton, M., Margon, B., Bowyer, S. 1976, \apj, 208, 177

\bibitem[Lanza (2008)]{Lanza08} 
Lanza, A.~F.\ 2008, \aap, 487, 1163 

\bibitem[Lanza (2010)]{Lanza10} 
Lanza, A.~F.\ 2010, \aap, 512, A77 

\bibitem[Lanza (2011)]{Lanza11} 
Lanza, A.~F.\ 2011, \apss, 336, 303 

\bibitem[Lanza et al. (2011a)]{Lanzaetal11a} 
Lanza, A.~F., Boisse, I., Bouchy, F., Bonomo, A.~S., \& Moutou, C.\ 2011a, \aap, 533, A44 

\bibitem[Lanza et al. (2010)]{Lanzaetal10} 
Lanza, A.~F., Bonomo, A.~S., Moutou, C., et al.\ 2010, \aap, 520, A53 

\bibitem[Lanza et al. (2011b)]{Lanzaetal11b} 
Lanza, A.~F., Bonomo, A.~S., Pagano, I., et al.\ 2011b, \aap, 525, A14 

\bibitem[\protect\citeauthoryear{Lanza et al.}{1998}]{Lanzaetal98}
Lanza, A. F., Catalano, S., Cutispoto, G., Pagano, I., Rodon\`o, M. 1998, \aap, 332, 541



\bibitem[\protect\citeauthoryear{Lanza et al.}{2009a}]{Lanzaetal09a}
Lanza, A. F., Pagano, I., G. Leto, S. Messina, S. Aigrain, et al. 2009a, \aap, 493, 193 

\bibitem[Lanza et al. (2009b)]{Lanzaetal09b} Lanza, A.~F., et al.\ 2009b, \aap, 506, 255 

\bibitem[\protect\citeauthoryear{Lanza et al.}{2003}]{Lanzaetal03}
Lanza, A. F., Rodon\`o, M., Pagano, I., Barge P., Llebaria, A. 2003, \aap, 403, 1135

\bibitem[\protect\citeauthoryear{Lanza et al.}{2004}]{Lanzaetal04}
Lanza, A. F., Rodon\`o, M., Pagano, I. 2004, \aap, 425, 707

\bibitem[\protect\citeauthoryear{Lanza et al.}{2007}]{Lanzaetal07}
Lanza, A. F., Bonomo, A. S., Rodon\`o, M. 2007, \aap, 464, 741


\bibitem[\protect\citeauthoryear{Lockwood et al.}{2007}]{Lockwoodetal07}
Lockwood, G. W., Skiff, B. A., Henry, G. W., Henry, S., Radick, R. R., Baliunas, S. L., Donahue, R. A., Soon, W. 2007, \apjs, 171, 260

\bibitem[Lou (2000)]{Lou00} 
Lou, Y.-Q.\ 2000, \apj, 540, 1102 


\bibitem[Massi et al. (2005)]{Massietal05} 
Massi, M., Neidh{\"o}fer, J., Carpentier, Y., \& Ros, E.\ 2005, \aap, 435, L1 


\bibitem[Mosser et al.(2009)]{Mosseretal09} 
Mosser, B., Baudin, F., Lanza, A.~F., Hulot, J.~C., Catala, C., Baglin, A., \& Auvergne, M.\ 2009, \aap, 506, 245 

\bibitem[Ogilvie \& Lin (2007)]{OgilvieLin07} 
Ogilvie, G.~I., \& Lin, D.~N.~C.\ 2007, \apj, 661, 1180 

\bibitem[Oliver et al. (1998)]{Oliveretal98} 
Oliver, R., Ballester,  J.~L., \& Baudin, F.\ 1998, \nat, 394, 552 

\bibitem[Pagano et al. (2009)]{Paganoetal09} 
Pagano, I., Lanza, A.~F., Leto, G., Messina, S., Barge, P., \& Baglin, A.\ 2009, Earth Moon and Planets, 105, 373 



\bibitem[\protect\citeauthoryear{Radick et al.}{1998}]{Radicketal98}
 Radick, R. R., Lockwood, G. W., Skiff, B. A., Baliunas, S. L. 1998, \apjs, 118, 239 


\bibitem[Rieger et al. (1984)]{Riegeretal84} 
Rieger, E., Kanbach, G., Reppin, C., et al.\ 1984, \nat, 312, 623 





\bibitem[Shkolnik et al.(2008)]{Shkolniketal08} 
Shkolnik, E., Bohlender, D.~A., Walker, G.~A.~H., \& Collier Cameron, A.\ 2008, \apj, 676, 628 

\bibitem[Silva-Valio \& Lanza (2011)]{SilvavalioLanza11} 
Silva-Valio, A., \& Lanza, A.~F.\ 2011, \aap, 529, A36 


\bibitem[Silva-Valio et al. (2010)]{Silvavalioetal10} 
Silva-Valio, A., Lanza, A.~F., Alonso, R., \& Barge, P.\ 2010, \aap, 510, A25 






\bibitem[Walker et al. (2007)]{Walkeretal07} Walker, G.~A.~H., Croll, 
B., Kuschnig, R., et al.\ 2007, \apj, 659, 1611 


\bibitem[Zaqarashvili et al. (2010)]{Zaqarashvilietal10} Zaqarashvili, 
T.~V., Carbonell, M., Oliver, R., \& Ballester, J.~L.\ 2010, \apj, 709, 749 


   
\end{thebibliography}
\end{document}